\documentclass[conference]{IEEEtran}
\IEEEoverridecommandlockouts
\usepackage{amsmath,amsfonts,amssymb}
\usepackage{graphicx}
\usepackage{cite}
\usepackage{cmap}
\usepackage{tabularx,ragged2e,booktabs}
\usepackage{caption}
\usepackage{subcaption}
\usepackage{hyperref}
\usepackage{tikz}
\usepackage{siunitx}
\usepackage{footnote}
\usepackage{multirow}
\usepackage{url}
\usepackage{algorithm}

\usetikzlibrary{shapes.multipart}
\makesavenoteenv{tabular}
\usepackage[left=0.625in, right=0.625in, top=0.75in, bottom=1.05in]{geometry}
\IEEEoverridecommandlockouts
\newcommand\copyrighttext{%
\footnotesize \textcopyright \enspace 2024 IEEE. Personal use of this material is permitted. Permission from IEEE must be obtained for all other uses, in any current or future media, including reprinting/republishing this material for advertising or promotional purposes, creating new collective works, for resale or redistribution to servers or lists, or reuse of any copyrighted component of this work in other works. DOI: \href{https://doi.org/10.1109/BlackSeaCom61746.2024.10646322}{10.1109/BlackSeaCom61746.2024.10646322}
}
\newcommand\copyrightnotice{%
\begin{tikzpicture}[remember picture,overlay]
\node[anchor=south] at (current page.south) {\fbox{\parbox{\dimexpr\textwidth-\fboxsep-\fboxrule\relax}{\copyrighttext}}};
\end{tikzpicture}%
}

\begin{document}

\title{Enhancing 5G V2X Mode 2 for Sporadic Traffic
\thanks{The research has been carried out at IITP RAS and supported by the Russian Science Foundation (Grant No 21-79-10431, https://rscf.ru/en/project/21-79-10431/)}
}

\author{\IEEEauthorblockN{Dmitry Bankov, Artem Krasilov, Artem Otmakhov, Aleksei Shashin, Evgeny Khorov} 
	
	\IEEEauthorblockN{Institute for Information Transmission Problems of the Russian Academy of Sciences, Moscow, Russia}
	
	\IEEEauthorblockN{Email: \{bankov, krasilov, otmakhov, shashin, khorov\}@wireless.iitp.ru}
}
\maketitle

\copyrightnotice
\begin{abstract}
	The emerging road safety and autonomous vehicle applications require timely and reliable data delivery between vehicles and between vehicles and infrastructure. To satisfy this demand, 3GPP develops a 5G Vehicle-to-Everything (V2X) technology. Depending on the served traffic type, 5G V2X specifications propose two channel access methods: (i) Mode 1, according to which a base station allocates resources to users, and (ii) Mode 2, according to which users autonomously select resources for their transmissions. In the paper, we consider a scenario with sporadic traffic, e.g., a vehicle generates a packet at a random time moment when it detects a dangerous situation, which imposes strict requirements on delay and reliability. To satisfy strict delay requirements, vehicles use Mode 2. We analyze the performance of Mode 2 for sporadic traffic and propose several approaches to improve it. Simulation results show that the proposed approaches can increase the system capacity by up to 40\% with a low impact on complexity. 
\end{abstract}


\section{Introduction}
\label{sec:intro}
5G Vehicle-to-Everything (V2X)~\cite{garcia2021tutorial} is a modern technology developed by 3GPP to provide mobile communications for autonomous vehicles and road safety applications, which are becoming more and more popular and widespread.
One of the key problems arising for the 5G V2X technology is satisfying the strict Quality of Service (QoS) requirements on the latency and reliability of data delivery set by the emerging applications.
The specific QoS requirements can be expressed in terms of delay budget $D^{QoS}$ and maximal packet loss rate $PLR^{QoS}$, which depend on the considered Level of Automation (LoA).
The requirements for different LoAs are presented in the 3GPP specifications~\cite{3gpp-v2x-scenarios}.
For example, low LoA scenarios require $D^{QoS} \sim \SI{100}{\ms}$ and $PLR^{QoS} \sim 10\%$.
In contrast, scenarios with a high LoA impose much stricter QoS requirements: $D^{QoS} \sim \SI{10}{\ms}$ and $PLR^{QoS} \sim 0.1\%$.

Specific QoS requirements in 5G V2X networks depend on the served traffic type.
Generally, vehicles (called UEs in 3GPP specifications) generate heterogeneous traffic that consists of periodic and sporadic flows.
For example, vehicles can periodically broadcast Cooperative Awareness Messages (CAMs), which contain information about their current position, velocity, acceleration, etc.
When a vehicle detects a critical situation on the road at a random time moment, it broadcasts a Decentralized Environmental Notification Message (DENM) to alert neighboring vehicles.
DENMs shall be delivered with low latency and high reliability because their loss can lead to dangerous consequences.

To serve such different kinds of traffic in vehicular scenarios, 5G V2X specifications support two channel access methods: Mode 1 and Mode 2.
According to Mode 1, a base station allocates resources for each UE transmission.
In contrast, Mode 2 allows UEs to send data autonomously using a variation of the multichannel slotted ALOHA method.
Mode 2 significantly reduces latency because it allows UEs to transmit data without waiting for a grant from the base station.

In this paper, we consider a 5G V2X network, where Mode 1 is used for sending CAMs and Mode 2 is used for sending DENMs, which allows satisfying strict latency requirements and reducing overhead associated with the exchange of control information with the base station.
We assume that non-overlapping frequency resources are allocated for serving CAMs and DENMs.
This assumption helps avoiding interference between different types of traffic and increases reliability~\cite{yin2022design,bankov2023analytical}.

Mode 2 allows UEs to transmit their packets several times achieving the required reliability.
Moreover, to decrease interference for retransmissions, UEs can use the resource reservation mechanism. According to this mechanism, using the control channel, a UE can inform the neighboring UEs about locations of further transmissions on the time-frequency grid.

To the best of our knowledge, the existing works studying the 5G V2X Mode 2 reservation mechanism either consider periodic traffic~\cite{sabeeh2019estimation,jeon2020explicit} or study sporadic traffic in scenarios with low LoA with moderate latency and reliability requirements~\cite{romeo2021supporting}.
In contrast to existing works, we consider the usage of 5G V2X Mode 2 for sporadic traffic in the high LoA scenarios.
Also, we propose several approaches to enhance the performance of 5G V2X Mode 2 and study their efficiency using system-level simulations.
As the key performance indicator, we consider the network capacity, which is defined as the maximum load that allows satisfying the given QoS requirements.
Simulation results show that the proposed approaches increase the capacity by up to 40\% in future V2X scenarios with strict latency and reliability requirements.

The rest of the paper is organized as follows.
Section~\ref{sec:works} analyses the related works.
Section~\ref{sec:channel} describes Mode 2, including retransmissions and resource reservation mechanism.
Section~\ref{sec:enhancements} describes the proposed Mode 2 enhancements.
Their performance evaluation is presented in Section~\ref{sec:performance}.
Finally, we draw conclusions in Section~\ref{sec:conclusion}.

\section{Related Works}
\label{sec:works}
5G V2X Mode 2 and a similar channel access method for LTE V2X networks (called C-V2X Mode 4) have been studied in many papers. In particular, papers~\cite{nba2020discrete, bankov2023analytical} develop analytical models to study the performance of V2X networks while serving the mixture of CAM and DENM traffic.
For example, the analytical model developed in~\cite{bankov2023analytical} allows estimating the network capacity and finding the optimal number of transmission attempts that maximizes the capacity.
However, these models do not consider the 5G V2X Mode 2 reservation mechanism and only model the random access procedure.

The authors of~\cite{he2018enhanced} propose a channel reservation mechanism that sends control messages over separate channel resources.
Apart from that, strategies for improving the reservation mechanism are proposed in~\cite{sabeeh2019estimation,jeon2020explicit}.
However, these papers only consider the LTE resource allocation mechanism that matches the periodic traffic and cannot be used for sporadic traffic.

Paper~\cite{yoon2021stochastic} proposes to improve the efficiency of the 5G V2X Mode 2 reservation mechanism by predicting the packet generation time and reserving the resources in advance.
Although such an approach shows a potential gain, the 3GPP specification~\cite{3gpp-phy} assumes that reservations are only used for retransmissions, but not for future transmission of new packets because resources can only be reserved up to 32 slots ahead.

Finally, the study~\cite{romeo2021supporting} investigates the impact of the reservation mechanism for retransmissions on the reliability of DENM delivery in the presence of interference caused by CAM transmissions.
The paper shows that the reservation mechanism is beneficial for increasing the reliability of DENM delivery. However, this study is limited only by a 100 ms delay budget and a reliability of 90\%, which corresponds to a low LoA scenario. Thus, the efficiency of the reservation mechanism for high LoA scenarios requires additional study.

Our paper aims to fill this research gap and studies the Mode 2 resource reservation mechanism with orthogonal resource allocation for CAMs and DENMs, as well as more strict requirements for latency and reliability.
We also propose several enhancements of Mode 2 and analyze their efficiency in scenarios with QoS requirements corresponding to low LoA and high LoA.

\section{5G V2X Mode 2 Overview}
\label{sec:channel}
\begin{figure}[!t]
	\centering
	\includegraphics[width=0.9\linewidth]{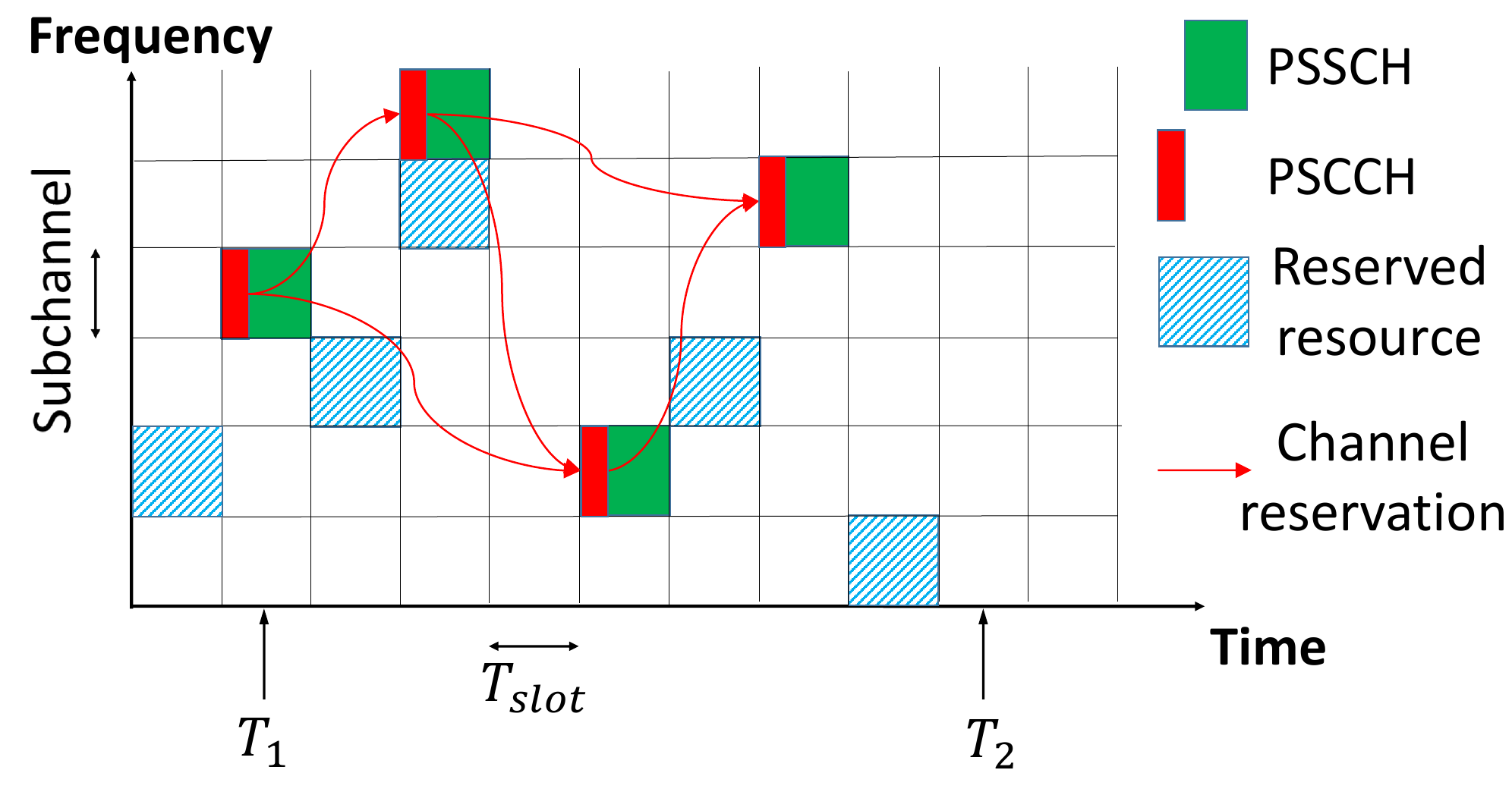}
	\caption{Time-frequency resources in Mode 2.}
	\label{fig:grid}
\end{figure}

According to 5G V2X Mode 2, UEs transmit data using the Orthogonal Frequency Division Multiple Access (OFDMA) method. Specifically, in the time domain, the channel resources are divided into slots of equal duration. In the frequency domain, the channel resources are split into subchannels of equal width (see Fig.~\ref{fig:grid}). For each transmission, a UE selects a slot and a continuous set of subchannels. The number of subchannels required for a packet transmission depends on the packet size, the subchannel width, and the used Modulation and Coding Scheme (MCS). Each transmission can be repeated several times to increase reliability.

At the physical layer, each transmission consists of (i) the Physical Sidelink Control Channel (PSCCH), which carries control data (i.e., the information about the transmission parameters and reservations of channel resources for future transmissions), (ii) the Physical Sidelink Shared Channel (PSSCH), which carries user data. PSCCH is encoded with the most robust MCS to provide the highest reliability, while the MCS for PSSCH can be adaptively selected based on the current channel conditions and traffic. The receiving UE shall first decode PSCCH. Then, based on the information obtained from PSCCH, it can decode PSSCH. If PSCCH is not decoded correctly, the UE misses the corresponding user data.

To select slots and subchannels for future transmissions, UEs use the sensing procedure, which works as follows. Each UE continuously senses the channel and decodes PSCCH. Each decoded PSCCH message contains the information about parameters of the current PSSCH transmission and about resources reserved for further transmissions: resources for retransmissions of the same packet and periodically reserved resources for transmission of the following packets. The 3GPP specifications allow reserving up to two resources for retransmissions corresponding to different time slots within a time window of 32 slots. However, each retransmission can provide a reservation for the next two retransmissions.

When a UE starts serving a new packet, it selects resources for $K$ transmissions (i.e., the initial transmission and $(K-1)$ retransmissions) according to the Standard Resource Selection algorithm (SRS), which works as follows. Depending on the packet delay budget and processing delay, the UE determines the set of slots (i.e., the interval $[T_1, T_2]$) that can be used for the packet transmission. Then, based on sensing information, the UE excludes the subchannels and the corresponding slots reserved by neighboring UEs. To avoid overbooking of channel resources by too far UEs, the 3GPP specifications introduce a special threshold $P_{th}$. If the average Reference Signal Received Power (RSRP) from the neighboring UE is higher than $P_{th}$, the reserved resources from this UE are considered busy. Otherwise, the reserved resources are marked as free. Having obtained the map of free resources, the UE randomly selects resources for $K$ transmissions. For each transmission attempt, the UE randomly selects a slot over the interval $[T_1, T_2]$. Inside the selected slot, the UE randomly selects such a starting subchannel that the transmission does not overlap with the reserved resources. If the UE does not find a suitable starting subchannel in the considered slot, it repeats the resource selection procedure by selecting another slot.

\section{Proposed Enhancements}
\label{sec:enhancements}

In this section, we propose several approaches aimed at improving the performance of 5G V2X Mode 2 for sporadic traffic with strict requirements on delay and reliability. In particular, the first approach modifies the SRS algorithm described in Section~\ref{sec:channel}. The second approach aims to improve the probability of delivering control information (including information about reserved resources) over PSCCH. The third approach is to apply full-duplex communication in 5G V2X systems. Let us consider them in more detail.

\subsection{Resource selection}
\label{subsec:first-left}
As described in Section~\ref{sec:channel}, the SRS algorithm selects slots randomly over the resource selection window $[T_1, T_2]$. Let us consider the case when UE A selects resources for $K$ transmission attempts. Let the first transmission attempt start on average $D$ slots after the beginning of the resource selection window. It can be shown~\cite{bankov2023analytical} that $D=\frac{T_2-T_1-K+1}{K+1}$. The average interval between consequent transmission attempts equals $(D+1)$ slots. The key drawback of the standard resource selection algorithm is that UE A delays the first transmission attempt, including the control information about the resources it has selected for subsequent attempts. Because of this delay, neighboring UEs are unaware of the future UE A's transmissions and can select overlapping resources with a high probability, which leads to collisions. 

To address this problem, we propose the Quick First Attempt (QFA) algorithm to change the way of selecting resources for the first transmission attempt.
With the QFA algorithm, the UE selects the first free slot for the first transmission attempt taking into account the map of already reserved resources. The resources for the second and subsequent attempts are selected following the SRS algorithm. On one side, the QFA algorithm significantly reduces the delay for transmitting control information, thus increasing the probability of successful transmission for subsequent attempts because other UEs become aware of these transmissions earlier. On the other side, the probability of a collision for the first transmission attempt increases. In Section~\ref{sec:performance}, we study the positive and negative effects of the QFA algorithm and compare it with the SRS algorithm.

\subsection{Sidelink control channel}
\label{subsec:reliable-ctrl}
Reliable delivery of PSCCH is important for providing high reliability of data transmission (i.e., successful PSSCH decoding). In particular, loss of control information leads to loss of user data for current transmission and information about locations of future transmissions and, therefore, a high probability of collisions for future transmissions.

On the transmitter side, PSCCH is encoded with the most robust MCS, which allows receiving it even when the signal power is much lower than noise and interference powers. On the receiver side, the probability of successful PSCCH reception significantly depends on the used decoding algorithm. The baseline decoding algorithm, which is further called Most Powerful PSCCH Decoding (MPPD), works as follows. When several UEs transmit PSCCH in the same subchannel, the receiving UE only tries to decode the PSCCH with the highest power. If the power of this signal is high enough with respect to other signals and noise, this signal is successfully decoded, while other signals are lost. The main advantages of MPPD are (i) low complexity, and (ii) low execution time because a single decoding iteration is used. However, in each subchannel, a UE cannot decode more than one PSCCH message 

In this paper, we study how far we can improve the system capacity if UEs implement an Ideal PSCCH Decoding (IPD) algorithm. With IPD, UEs can successfully decode all interfering PSCCH signals. In practice, such kind of algorithms can be implemented based on 
Successive Interference Cancellation (SIC) technique~\cite{sic_fundamental}. With SIC, interfering signals are decoded iteratively. A UE first tries to decode the signal with the highest power. If decoding is successful, the UE reconstructs the decoded signal and subtracts it from the sum of all signals. Then, the UE considers the next signal, etc. IPD can be considered as the upper bound for the SIC-based decoders when a UE can decode all signals. Note that SIC-based decoders have much higher complexity and execution time. Therefore, in Section~\ref{sec:performance}, we analyze the benefits of such a solution.

\subsection{Full-duplex communication}
\label{subsec:full-duplex}
According to 3GPP Release 16 specifications, V2X UEs are assumed to work in Half-Duplex (HD) mode. HD means that a UE cannot simultaneously transmit and receive data in the same slot. The reason is that the power of the generated signal is several orders of magnitude higher than the power of the received signal. 

Recently published 3GPP Release 18 specifications~\cite{3gpp-sbfd} enabled Subband non-overlapping Full-Duplex (SBFD). SBFD is originally designed for the case when UEs and a base station transmit data in the same slot but use different subbands (subchannels). For example, a base station can simultaneously transmit data to some UE in one subchannel and receive data from another UE in another subchannel. Since transmission and reception occur at different frequencies the self-interference (i.e., interference of the transmitting signal to the received one) can be significantly reduced. In the paper, we propose to apply  SBFD for V2X scenarios. In particular, we assume that a V2X UE transmitting in some subchannel can simultaneously receive data in other subchannels. To estimate the upper bound of the gain provided by SBFD, we assume zero self-interference between different subchannels (i.e., the signal generated in one subchannel creates zero interference in other subchannels).

Currently, 3GPP and the researchers are working on the design of In-band Full-Duplex (IBFD)\cite{smida2023full}, which is expected to be a part of 6G systems. With IBFD, a base station or a UE can simultaneously transmit and receive data in the same slot and the same subchannel. The key research issue for IBFD is the design of robust self-interference cancellation methods. In the paper, we analyze the upper bound that can be provided by IBFD, assuming that UEs implement ideal self-interference cancellation. 

\section{Performance Evaluation}
\label{sec:performance}

\subsection{Simulation setup}
\label{subsec:simulation}

To analyze the performance of Mode 2 and evaluate the approaches proposed in Section~\ref{sec:enhancements}, we use the NS-3 network simulator~\cite{ns3} with the V2X model developed in~\cite{cttc}. We extended this model to evaluate scenarios with sporadic traffic and implemented the approaches presented in Section~\ref{sec:enhancements}.

We consider a scenario with $N$ UEs moving on a highway. Following~\cite{todisco2021performance}, the distance between neighboring UEs has an exponential distribution with the average distance $r=10$ m.
Each UE generates 300-byte DENM packets and broadcasts them using MCS 6~\cite{romeo2021supporting}. A packet shall be delivered to all UEs within a circle of radius $R$ (so-called relevance area).
We consider a log-distance propagation model: $L(d)=L_0+10 \cdot n \cdot \log_{10}(\frac{d}{d_0})$, where $L(d)$ is the path loss in dB for two UEs located at distance $d$.
The main simulation parameters are presented in Table~\ref{tab:parameters}.

\begin{table}[!t]
	\caption{Main simulation parameters.}
	\label{tab:parameters}
	\begin{center}
		\begin{tabular}{|l|c|}
			\hline
			\textbf{Parameter} & \textbf{Value} \\
			\hline
			Number of UEs $N$ & 200 \\
			Transmission power $S$ & 23 dBm \\
			Noise figure & 5 dB \\
			Slot duration & $500$ $\mu$s \\
			Average distance between UEs $r$ & 10 m \\
			Packet size & 300 bytes \\
			MCS for PSSCH & 6 \\
			Reference distance $d_0$ & 1 m \\
			Reference loss at reference distance $L_0$ & 46,7 dB\\
			Path loss exponent $n$ & 3 \\
			Radius of the relevance area $R$ & 200 m \\
			\hline
		\end{tabular}
	\end{center}
\end{table}

We consider two reference sets of QoS requirements: (i) low LoA with $D^{QoS}=10$~ms and $PLR^{QoS}=0.1$, (ii) high LoA with $D^{QoS}=10$~ms and $PLR^{QoS}=10^{-3}$.

We evaluate the network performance using the following indicators.
In each simulation run, we measure the packet loss rate $PLR_i$ of UE $i$ as the average fraction of UEs within a relevance area that do not receive the DENM packet successfully. PLR for the entire network $PLR$ is estimated by averaging $PLR_i$ across all UEs.
The network capacity $C$ is defined as the maximum load for which $PLR\leq PLR^{QoS}$.

\subsection{Results analysis}

\begin{figure}[!t]
	\centering
	\includegraphics[width=0.9\linewidth]{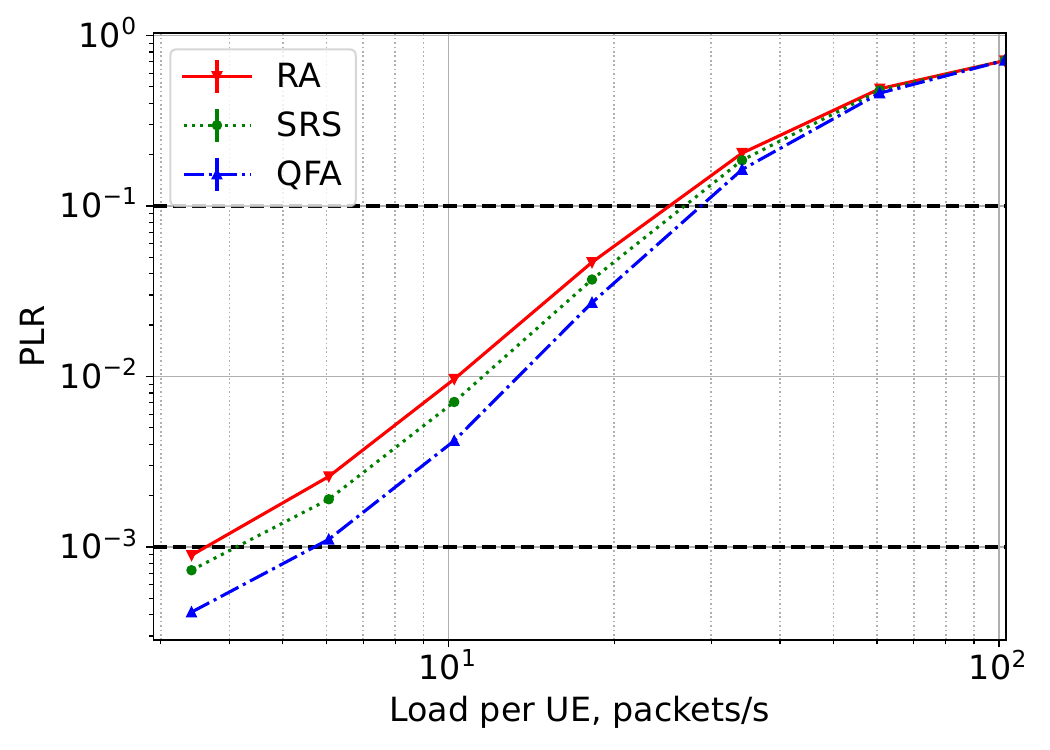}
	\caption{PLR as the function of load, $K=5$.}
	\label{fig:plr}
\end{figure}

\begin{figure}[!t]
	\centering
	\includegraphics[width=0.9\linewidth]{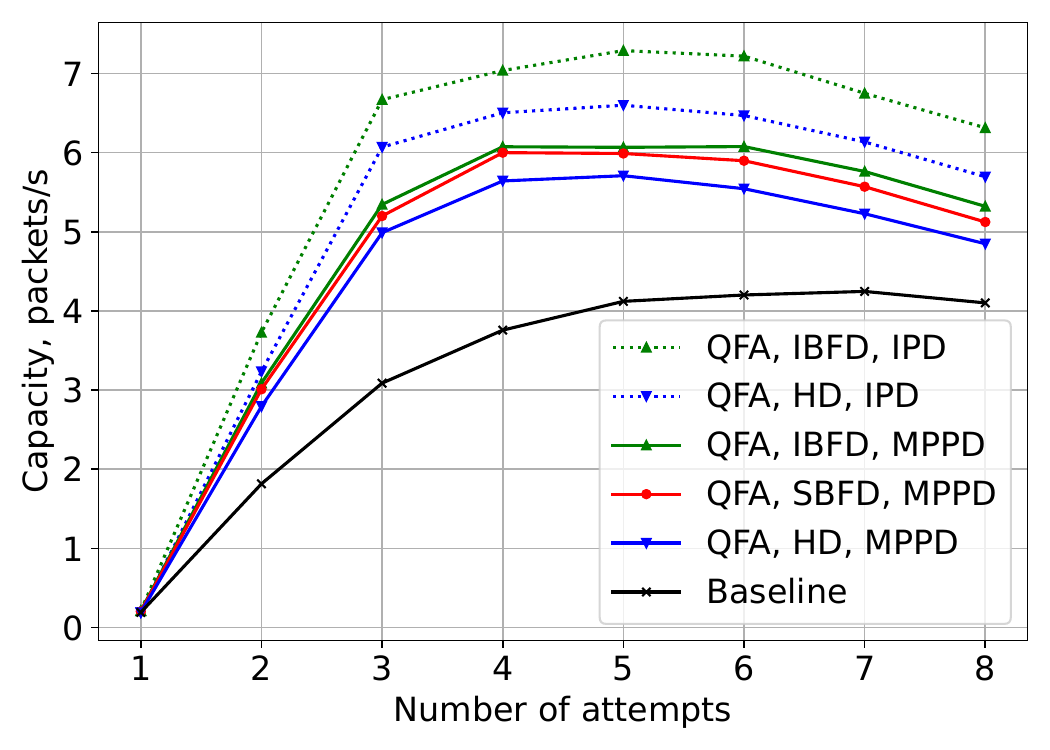}
	\caption{Capacity for high LoA scenario ($PLR^{QoS}=10^{-3}$).}
	\label{fig:cap_1e-3}
\end{figure}

Let us first analyze the performance of various resource selection algorithms described in Section~\ref{subsec:first-left}. Figure~\ref{fig:plr} shows PLR as the function of load when all UEs use $K=5$ transmission attempts. As we show further (see Fig.~\ref{fig:cap_1e-3}), $K=5$ provides the maximum capacity for the case of strict reliability requirement $PLR^{QoS}=10^{-3}$. We compare three resource selection algorithms: (i) the Random Access (RA) algorithm that does not use the information about the reserved resources, (ii) the SRS algorithm that takes into account the reservations and selects slots randomly in the resource selection window, and (iii) the proposed QFA algorithm that selects the earliest possible slot for the first transmission attempt. 

We can see that the SRS algorithm provides only slight improvement (up to 10\%) of capacity for any selected PLR level. Thus, the knowledge of additional information about future transmissions combined with the SRS algorithm insignificantly improves the system performance. In contrast, the proposed QFA algorithm noticeably increases the capacity: (i) by 15\% for $PLR^{QoS}=0.1$, and (ii) by 40\% for $PLR^{QoS}=10^{-3}$ with respect to the SRS algorithm. We can see that the efficiency of the QFA algorithm significantly increases for future scenarios with a high reliability requirement. The reason for this effect is as follows. High reliability is achieved at low load, which means that only few resources are occupied in the resource selection window. Thus, the QFA algorithm has more chances to select the first available slot and provide information about reserved resources right after packet arrival. The latter reduces the probability of collisions for retransmissions and, in turn, improves data delivery reliability.

\begin{figure}[!t]
	\centering
	\includegraphics[width=0.9\linewidth]{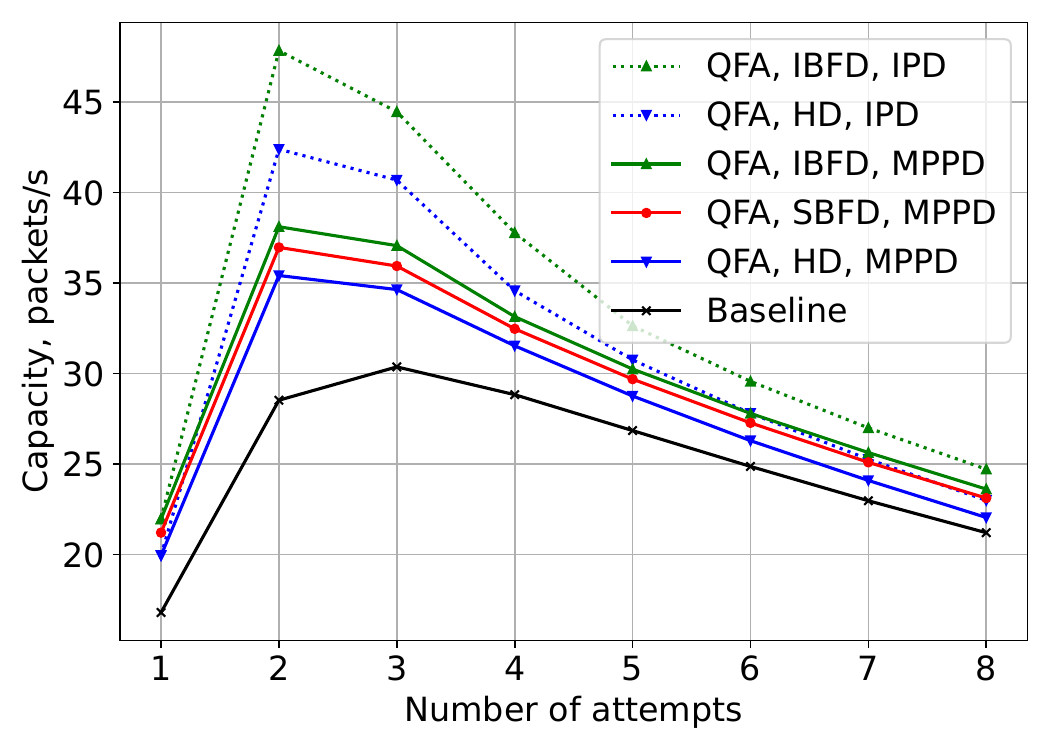}
	\caption{Capacity for low LoA scenario ($PLR^{QoS}=0.1$).}
	\label{fig:cap_1e-1}
\end{figure}

Let us study the dependency of the network capacity on the number of packet transmission attempts.
Figure~\ref{fig:cap_1e-1} and Figure~\ref{fig:cap_1e-3} show this dependency for $PLR^{QoS}=0.1$ and $PLR^{QoS}=10^{-3}$, respectively.
The ``Baseline'' curve shows the results for the SRS algorithm, MPPD decoding, and HD mode.
Other curves show the results of the QFA algorithm used with different PSCCH decoding algorithms and duplex schemes.
First, we see that the usage of the QFA algorithm significantly increases the capacity with respect to the SRS algorithm.
Second, we see that the optimal number of transmission attempts depends on the PLR requirement: (i) for $PLR^{QoS}=0.1$ we should use $K=2$, (ii) for $PLR^{QoS}=10^{-3}$ we should use $K=5$ for all the solutions that use QFA.
In addition, with low LoA, the capacity quickly degrades when $K$ becomes greater than the optimal value, while for high LoA, this degradation is slower.
The reason is that high reliability is achieved at low load. Thus, UEs can use free resources to perform more transmission attempts. However, too high number of transmission attempts increases the collision probability. The interested reader can refer to our recent paper~\cite{bankov2023analytical}, which provides an analytical model for estimating the optimal $K$.

Let us study the influence of the PSCCH decoding algorithm.
On Fig.~\ref{fig:cap_1e-1} and Fig.~\ref{fig:cap_1e-3} we see that for optimal $K$, IPD algorithm increases the capacity (i) by up to 20\% for $PLR^{QoS}=0.1$, and (ii) by up to 15\% for $PLR^{QoS}=10^{-3}$ compared with the MPPD algorithm. Thus, even the usage of the ideal PSCCH decoding algorithm provides a slight improvement in data delivery reliability. 

Let us study the added value of the full-duplex schemes described in Section~\ref{subsec:full-duplex}. Figure~\ref{fig:cap_1e-1} and Figure~\ref{fig:cap_1e-3} show that implementation of either SBFD or IBFD with the baseline MPPD decoding algorithm provides a tiny capacity improvement that does not exceed 5\%. Note that, as expected, IBFD provides a higher capacity than SBFD. Only the usage of the advanced PSCCH decoding algorithm (i.e., the IPD algorithm) together with the IBFD scheme provides a notable gain (i) of up to 35\% for $PLR^{QoS}=0.1$, and (ii) of up to 25\% for $PLR^{QoS}=10^{-3}$ with respect to HD and MPPD. However, we note that this gain corresponds to the upper bound that can be achieved with an ideal implementation of the PSCCH decoding algorithm and full-duplex scheme. Moreover, the gain comes at the cost of a significant increase in the V2X UE complexity and requires modification of V2X UE hardware. So, designing low-complexity algorithms that can use most of the gain requires additional study. 

\section{Conclusion}
\label{sec:conclusion}

In this paper, we have proposed and analyzed various approaches to improve the performance of 5G V2X Mode 2 for sporadic traffic. In particular, the first approach modifies the algorithm for selecting channel resources for the initial transmission and can be easily implemented on the existing hardware. Simulation results show that the proposed approach increases the system capacity by up to 40\% in future V2X scenarios with a high-reliability requirement. The second approach is to implement the advanced PSCCH decoding algorithm that allows increasing the probability of delivering control information based on successive interference cancellation techniques. We show that this approach provides additionally up to 20\% gain. The third approach is to apply various full-duplex schemes recently developed by 3GPP and the research community. Our results show that full-duplex schemes provide notable gain (up to 35\%) only when they are implemented together with the advanced PSCCH decoding algorithms. However, the implementation of full-duplex schemes requires modification of the V2X UE hardware and requires the design of low-complexity algorithms~\cite{stanojevic2023convolutional, trifonov2023design} that can be considered as the direction for future study.

\bibliographystyle{IEEEtran}
\bibliography{biblio}

\end{document}